\documentclass[10pt,conference]{IEEEtran}

\usepackage{amsmath}
\usepackage{amssymb}
\usepackage{amsfonts}
\usepackage{graphicx}
\usepackage{graphics}
\usepackage{theorem}
\usepackage{euscript}
\usepackage{psfrag}

\def\E{{\mathbb E}}        
\def\1{{\mathbf 1}}        

\theoremstyle{plain}

\newcommand{\ebno}{{E_b/N_0}}

\newcommand{\febno}{\frac{E_b}{N_0}}

\newcommand{\So}{\mathcal{S}_0}
\newcommand{\Sinf}{\mathcal{S}_\infty}
\newcommand{\Linf}{\mathcal{L}_\infty}

\newcommand{\mysize}{\small}

\newcommand{\vct}[1]{\boldsymbol{#1}}
\newcommand{\Mat}[1]{\boldsymbol{#1}}
\newcommand{\abs}[1]{\left\vert#1\right\vert}

\DeclareMathOperator{\trace}{tr}

\begin{document}

\title{On the Spectrum of Large Random Hermitian Finite-Band Matrices}

\author{\authorblockN{Oren Somekh\authorrefmark{1}, Osvaldo Simeone\authorrefmark{2}, Benjamin M. Zaidel\authorrefmark{3},
H. Vincent Poor\authorrefmark{1}, and Shlomo Shamai
(Shitz)\authorrefmark{4}}
\authorblockA{\authorrefmark{1}
Department of Electrical Engineering, Princeton University,
Princeton, NJ 08544, USA}
\authorblockA{\authorrefmark{2}
CWCSPR, Department of Electrical and Computer Engineering, NJIT,
Newark, NJ 07102, USA}
\authorblockA{\authorrefmark{3}
Department of Electronics and Telecommunications, NTNU, Trondheim
7491, Norway}
\authorblockA{\authorrefmark{4}
Department of Electrical Engineering, Technion, Haifa 32000,
Israel}}

\maketitle

\begin{abstract}
The open problem of calculating the limiting spectrum (or its
Shannon transform) of increasingly large random Hermitian
finite-band matrices is described. In general, these matrices
include a finite number of non-zero diagonals around their main
diagonal regardless of their size. Two different communication
setups which may be modeled using such matrices are presented: a
simple cellular uplink channel, and a time varying inter-symbol
interference channel. Selected recent information-theoretic works
dealing directly with such channels are reviewed. Finally, several
characteristics of the still unknown limiting spectrum of such
matrices are listed, and some reflections are touched upon.
\end{abstract}

\section{Problem Description}
Consider a linear channel of the form
\begin{equation}\mysize\label{eq: Received signal}
    \vct{y}=\Mat{H}_N\vct{x}+\vct{z}\ ,
\end{equation}
where $\vct{x}$ is the $NK\times 1$ zero-mean complex Gaussian
input vector $\vct{x}\sim \mathcal{CN}(0,\frac{P}{K}\Mat{I}_{NK})$
\footnote{An $N\times N$ identity matrix is denoted by
$\Mat{I}_N$.}, $\vct{y}$ is the $N\times 1$ output vector, and
$\vct{z}$ denotes the $N\times 1$ zero-mean complex Gaussian
additive noise vector $\vct{z}\sim \mathcal{CN}(0,\Mat{I}_{NK})$,
which is independent of $\vct{x}$ and $\Mat{H}_N$. Accordingly
$\rho=\frac{P}{K}$ is the transmitted signal-to-noise ratio (SNR).
In addition, the $N\times NK$ channels transfer matrix $\Mat{H}_N$
is defined by
\begin{equation}\mysize\label{eq: Wyner transfer matrix}
\Mat{H}_N=\left(
\begin{array}{cccccc}
\vct{a}_{1} & \beta\vct{c}_{1} & \vct{0} & \cdots & \vct{0} & \vct{0} \\
\alpha\vct{b}_{2} & \vct{a}_{2} & \beta\vct{c}_{2} & \vct{0} & \cdots & \vct{0} \\
\vct{0} & \alpha\vct{b}_{3} & \vct{a}_{3} & \beta\vct{c}_{3} & \ddots & \vdots \\
\vdots & \vct{0} & \alpha\vct{b}_{4} & \ddots & \ddots & \vct{0} \\
\vct{0} & \vdots & \ddots & \ddots & \vct{a}_{N-1} & \beta\vct{c}_{N-1} \\
\vct{0} & \vct{0} & \cdots & \vct{0} & \alpha\vct{b}_{N} &
\vct{a}_{N}
\end{array}
\right) \ ,
\end{equation}
where $\{\vct{a}_{i},\ \vct{b}_{i},\ \vct{c}_{i}\}$ are
statistically independent $1\times K$ random row vectors with
independent identically distributed (i.i.d.) entries
$a_{i,j}\sim\pi_a$, $b_{i,j}\sim\pi_b$, and $c_{i,j}\sim\pi_c$.
For simplicity, we assume that the power moments of the entries
for any finite order are bounded.
Finally, $\alpha,\ \beta\in[0,1]$ are constants.

The normalized input-output mutual information of \eqref{eq:
Received signal} conditioned on $\Mat{H}_N$ (also known as the
Shannon transform) is\footnote{Unless explicitly denoted otherwise
a natural base logarithm is used throughout this presentation.}
\begin{equation}\mysize\label{eq: Mutual information}
\begin{aligned}
    \frac{1}{N}I(\vct{x};\vct{y}|\Mat{H}_N)&=\frac{1}{N}\log\det\left(\Mat{I}_N+\rho\Mat{H}_N\Mat{H}_N^\dagger\right)\\
    &=\frac{1}{N}\sum_{i=1}^N\log\left(1+\rho\lambda_i(\Mat{H}_N\Mat{H}_N^\dagger)\right)\\
    &=\int_0^\infty\log(1+\rho
    x)d\mathrm{F}_{\Mat{H}_N\Mat{H}_N^\dagger}(x)\ ,
\end{aligned}
\end{equation}
where $\lambda_i(\Mat{H}_N\Mat{H}_N^\dagger)$ denotes the $i$th
eigenvalue of the Hermitian \emph{five-diagonal} matrix
$\Mat{H}_N\Mat{H}_N^\dagger$. Furthermore, denoting the indicator
function by $1\{\cdot\}$,
\begin{equation}\mysize\label{eq: Spectrum}
\mathrm{F}_{\Mat{H}_N\Mat{H}_N^\dagger}(x)=\frac{1}{N}\sum_{i=1}^N\vct{1}\{\lambda_i(\Mat{H}_N\Mat{H}_N^\dagger)\le
x\}
\end{equation}
is the empirical cumulative distribution function of the
eigenvalues (also referred to as the spectrum or empirical
distribution) of $\Mat{H}_N\Mat{H}_N^\dagger$. Fixing $K$ and
assuming that $\mathrm{F}_{\Mat{H}_N\Mat{H}_N^\dagger}(x)$
converges almost surely (a.s.) to a unique limiting spectrum
$\mathrm{F}_{\Mat{H}_N\Mat{H}_N^\dagger}(x)\underset{N\rightarrow\infty}{\overset{\mathrm{a.s.}}\longrightarrow}F(x)$,
it can be shown that the expectation of \eqref{eq: Mutual
information} with respect to (w.r.t.) the distribution of
$\Mat{H}_N$ converges as well. This is since \eqref{eq: Mutual
information} is uniformly integrable due to the Hadamard
inequality and the bounded power moment assumption, and hence the
a.s. convergence implies convergence in expectation
\cite{Levy-Somekh-Shamai-Zeitouni-IT07}.

In Section \ref{sec: Motivation} it will be realized that if the
channel $\Mat{H}_N$ is known at the receiver and its variation
over time is stationary and ergodic, then the expectation of
\eqref{eq: Mutual information} w.r.t. the distribution of
$\Mat{H}_N$ is the per-cell sum-rate capacity of a certain
cellular uplink channel model. In another setting (see Section
\ref{sec: Motivation}), the same expectation may be interpreted as
the capacity of a certain time variant inter-symbol interference
(ISI) channel, assuming again that the channel is known at the
receiver.

\subsection{Analytical Difficulty}\label{sec: Analysis Difficulty}

Many recent studies have analyzed the asymptotic rates of various
vector channels using results from the theory of (large) random
matrix (see \cite{Tulino-Verdu-Random-Matrix-Review-2004} for a
recent review). In those cases, the number of random variables
involved is of the order of the number of elements in the matrix
$\Mat{H}_N$, and self-averaging is strong enough to ensure
convergence of the empirical measure of eigenvalues, and to derive
equations for the limiting spectrum (or its Stieltjes transform).
In particular, this is the case if the normalized continuous power
profile of $\Mat{H}_N$, which is defined with $r,\ t\in[0,1]$ as
\begin{equation}\mysize\label{eq: MIMO definition of the power profile}
\begin{aligned}
\mathcal{P}_N(r,t)&\triangleq \E(\abs{[\Mat{H}_N]_{i,j}}^2) \\
&\frac{i-1}{N} \le r < \frac{i}{N}\ , \ \frac{j-1}{NK} \le t <
\frac{j}{NK}\ ,
\end{aligned}
\end{equation}
converges uniformly to a bounded, piecewise continuous function as
$N\to\infty$, see e.g. \cite[Theorem
2.50]{Tulino-Verdu-Random-Matrix-Review-2004}.
In the case
under consideration here, it is easy to verify that for $K$ fixed,
$\mathcal{P}_M(r,t)$ does \emph{not} converge uniformly, and other
techniques are required.

\noindent \textbf{Remark:} It is noted that the setting of
\eqref{eq: Received signal} can be extended in many ways such as
increasing the number of non-zero block diagonals, or replacing
each $K$-dimensional random row vector with an $n\times m$ random
matrix. Such settings result in $\Mat{H}_N\Mat{H}_N^\dagger$ which
includes more than five non-zero diagonals and are referred to as
Hermitian \emph{finite-band} random matrices; the resulting
matrices contain only zero entries outside a finite band (finite
number of non-zero diagonals) around their main diagonal
regardless of $N$.

To the best of the authors' knowledge, neither the limiting
spectrum of Hermitian finite-band random matrices, nor the
expectation of the normalized input-output conditional mutual
information \eqref{eq: Mutual information}, is known in general
except for a few special cases (see Section \ref{sec: Selected
Prior Work}). Moreover, even the high-SNR regime characterization
(defined in
\cite{Shamai-Verdu-2001-fading}\cite{Lozano-Tulino-Verdu-high-SNR-IT05})
of the latter is known only for a few special cases (see Section
\ref{sec: Selected Prior Work}) and remains an open problem in
general.

\section{Motivation}\label{sec: Motivation}

In this section we present two different multi-access
communication channels whose channel transfer matrices are
finite-band.

\paragraph{Cellular uplink} Motivated by the fact that a mobile user in a cellular
system effectively ``sees" only a finite number of base-stations,
a simplified cellular model family has been introduced by Wyner in
\cite{Wyner-94} (see also \cite{Hanly-Whiting-Telc-1993} for an
independent earlier work which deals with similar setups).
According to the original linear variant setup presented in
\cite{Wyner-94}, the $K$ homogenous users of each cell are
collocated at the cell's center and ``see" their local
base-station antenna and the antennas of the two adjacent
base-stations only. While the signals travel to the local antenna
with no path-loss, the path-losses to the adjacent cell antenna on
the left and to the one on the right, are characterized by two
parameters $\alpha,\beta\in[0,\ 1]$, respectively. Wyner assumed
that the users cannot cooperate in any way and that all the
base-stations are connected to a central receiver via an ideal
error-free infinite capacity backhaul network. With optimal joint
processing of all the received signals, the channel can be
considered as a multiple-access channel whose vector
representation is given by \eqref{eq: Received signal}. The
non-fading setup of \cite{Wyner-94} was extended to include flat
fading channels in
\cite{Shamai-Wyner-97-I-II}\cite{Somekh-Shamai-2000}. Considering
an infinite number of cells and assuming that the channel state
information is known by the central receiver, the per-cell
sum-rate capacity of the Wyner model is given by setting
$\pi_a=\pi_b=\pi_c=\pi$, and averaging the mutual information of
\eqref{eq: Mutual information} over the entries of $\Mat{H}_N$. It
is noted that the basic model can be extended to cases where each
mobile ``sees" any finite number of cell-site antennas and the
resulting $\Mat{H}_N\Mat{H}_N^\dagger$
is a finite-band matrix.\\
\noindent \textbf{Remark:} Using the uplink-downlink duality (e.g.
\cite{Viswanath-Tse-2003}), the per-cell sum-rate capacity of the
Wyner uplink channel is also an achievable per-cell sum-rate (a
lower bound of the per-cell capacity) of the reciprocal Wyner
downlink channel, assuming the joint multicell transmitter has
full channel state information (CSI) while each mobile is aware of
its own CSI only.

Since its introduction in \cite{Wyner-94}, the Wyner model family
has provided a powerful framework for research assessing the
performance of various joint multicell processing schemes (see
\cite{Somekh-Simeone-Barness-Haimovich-Shamai-BookChapt-07} and
\cite{Shamai-Somekh-Zaidel-JWCC-2004} for recent surveys).
Overcoming the analytical difficulties relating to these models
and calculating the spectra (or their transforms) of the resulting
finite-band matrices, would greatly enhance our understanding and
insight into the theoretical performance of future cellular (and
wireless) systems.

\paragraph{Time varying ISI channels} Here we consider $K$ homogenous users
communicating with a receiver over an $L$-tap time varying ISI
channel. Assuming that the channel taps are i.i.d. between
different users and also i.i.d. in the time index it is easily
verified that the received signal is given by \eqref{eq: Received
signal}. Assuming that $L=3$, the sum-rate of this multiple access
channel is given by averaging the mutual information of \eqref{eq:
Mutual information} over the entries of $\Mat{H}_N$. This setup
may describe a ``fast" multipath fading channel where the channel
taps are independent over the time index. As with the previous
setup for any finite $L$ the resulting
$\Mat{H}_N\Mat{H}_N^\dagger$ is a finite-band matrix. In contrast
to the previous model where the entries of the received signal are
in the spatial domain, the entries of the received signal here are
in the time domain.

\section{Selected Prior Work}\label{sec: Selected Prior Work}

In this section we briefly review selected previous works dealing
with the spectrum of finite-band matrices, its Shannon transform,
and related issues. The reader is referred to
\cite{Somekh-Simeone-Barness-Haimovich-Shamai-BookChapt-07} and
\cite{Shamai-Somekh-Zaidel-JWCC-2004} for detailed surveys of
relevant information-theoretic works.

The non-fading (or deterministic) case was analyzed by Wyner in
\cite{Wyner-94} for the special case of $\beta=\alpha$. Setting
$a_{i,j}=b_{i,j}=c_{i,j}=1$ we get that
$\frac{1}{K}\Mat{H}_N\Mat{H}_N^\dagger$ becomes a five-diagonal
Toeplitz matrix with non-zero entries $(\alpha^2,\ 2\alpha,\
1+2\alpha^2,\ 2\alpha,\ \alpha^2)$. Using well known results
regarding the limiting spectrum of large Toeplitz matrices
(Szeg{\"o}'s Theorem \cite{Gray-paper-72}), Wyner showed that the
per-cell sum-rate capacity approaches as $N\rightarrow\infty$ to
\begin{equation}\mysize\label{eq: Wyner rate}
    C=\int_0^1\log\left(1+P(1+2\alpha\cos(2\pi f))^2\right)df\ .
\end{equation}
It is noted that the result is independent of $K$ as long as the
total transmit power per-cell $P$ is fixed. The reader is referred
to \cite{Letzepis-PhD06} for a derivation of the Stieltjes
transform of the spectrum for similar five-diagonal Toeplitz
matrices.

The infinite linear Wyner model in the presence of flat fading
channels is considered in \cite{Somekh-Shamai-2000}. For the
special case of $\beta=\alpha$, $\pi_a=\pi_b=\pi_c=\pi$ and $K=1$
it is shown that the unordered eigenvalue distribution
$\mysize\E(F_{\Mat{H}_N\Mat{H}_N^\dagger})$ converges weakly to a
unique distribution. It is conjectured that using similar methods
the spectrum can be proved to converge a.s. to a unique limit as
well. In addition, using a standard weighted paths summation over
a restricted grid, the limiting values of the first several
moments of this distribution were calculated for the special case
in which the amplitude of an individual fading coefficient is
statistically independent of its uniformly distributed phase (e.g.
\emph{Rayleigh} fading $\pi=\mathcal{CN}(0,1)$). For example,
listed below are the first three limiting moments:
\begin{equation}\mysize\label{eq: Wyner limit moments}
\begin{aligned}
\mathcal{M}_1=&m_2 + 2 m_2 \alpha ^2\\
\mathcal{M}_2=&m_4 + 8 m_2^2\alpha^2 + (4m_2^2 +
2m_4)\alpha^4\\
\mathcal{M}_3=&m_6 + (6m_2^3 + 12m_2m_4)\alpha^2 + (36m_2^3 +
12m_2m_4)\alpha^4\\
&\quad\quad\quad\quad+(6m_2^3 + 12m_2m_4 + 2m_6)\alpha^6\ ,
\end{aligned}
\end{equation}
where $m_{i}$ is the $i$-th power moments of the amplitude of an
individual fading coefficient. It is noted that this procedure can
be extended in principle, although in a tedious manner, for any
finite $K$ or also for $\Mat{H}_N$ to include more than three
non-zero block diagonals. Since the limiting moments of increasing
order are functions of increasing orders of the moments of the
fading coefficients, it is conjectured that the limiting
distributions (and also the spectra) of finite-band matrices
depend on the \emph{actual} fading distribution and not just on
its few first moments.
Focusing on the case in which $K$ is large while $P$ is kept
constant, and applying the strong law of large numbers (SLLN), the
entries of $\frac{1}{K}\boldsymbol{H}_N\boldsymbol{H}_N^{\dagger
}$ consolidate a.s. to their mean values and the latter becomes a
Toeplitz matrix. By applying Szeg{\"o}'s Theorem for
$N\rightarrow\infty$ it is shown in \cite{Somekh-Shamai-2000} that
the per-cell sum-rate capacity is given by
\begin{multline}{\mysize\label{eq: Wyner-lin opt
fading large K} C=\int_{0}^{1}\log \left( 1+P\left[\right.\right.
\sigma^2(1+2\alpha ^{2})}\\
{\mysize\left.\left.+\abs{m_1}^2\left( 1+2\alpha \cos (2\pi \theta
)\right) ^{2}\right] \right) d\theta}\ ,
\end{multline}
where $\sigma^{2}=m_2-\abs{m_1}^2$ is the variance of an
individual fading coefficient.

An alternative approach which replaces the role of the eigenvalues
of $\boldsymbol{H}_N\boldsymbol{H}_N^{\dagger }$ with the diagonal
elements of its \emph{Cholesky} decomposition, is presented by
Narula \cite{Narula-1997}. With $\alpha=1$, $\beta=0$,
$\pi_a=\pi_b=\pi$, and $K=1$, the resulting
$\boldsymbol{H}_N\boldsymbol{H}_N^{\dagger }$ is a three-diagonal
matrix (also known as \emph{Jacobi} matrix). Originally, Narula
has studied the capacity of a ``fast" time varying two-tap ISI
channel, where the channel coefficients are i.i.d. zero-mean
complex Gaussian (i.e. $\pi=\mathcal{CN}(0,1)$). Following
\cite{Narula-1997}, the diagonal entries of the \emph{Cholesky}
decomposition applied to the covariance matrix
$\mysize{\left(\boldsymbol{I}_N+P\boldsymbol{H}_N\boldsymbol{H}_N^{\dag
}\right) =\boldsymbol{L}_N\boldsymbol{D}_N\boldsymbol{U}_N}$, are
given by
\begin{equation}\mysize\label{eq: RM Narula
Cholesky} d_{n}=1+P\left\vert a_{n}\right\vert ^{2}+P\left\vert
b_{n}\right\vert ^{2}\left( 1-P\frac{\left\vert a_{n-1}\right\vert ^{2}%
}{d_{n-1}}\right) \ ,\ n=2,\ldots ,N\ ,
\end{equation}
with an initial condition $d_{1}=1+P\left\vert a_{1}\right\vert
^{2}+P \left\vert b_{1}\right\vert ^{2}$. Thus, the  diagonal
entries $\{d_{m}\}$ form a discrete-time continuous space Markov
chain. Remarkably, Narula managed to prove that this Markov chain
possesses a unique ergodic stationary distribution, given by
\begin{equation}\mysize
f_{d}(x)=\frac{\log (x)e^{-\frac{x}{\bar{P}}}}{\text{Ei}\left( \frac{1}{\bar{%
P}}\right) \bar{P}}\quad ;\quad x\geq 1\ , \label{eq: RM Narula
stationary distribution}
\end{equation}
where $\text{Ei}(x)=\int_{x}^{\infty }\frac{\exp (-t)}{t}dt$ is
the exponential integral function. Further, it is proven in
\cite{Narula-1997} that the SLLN holds for the sequence $\{\log
{d_{n}}\}$ as $N\rightarrow \infty$, and the channel capacity is
\begin{equation}\mysize\label{eq: RM TDMA uplink capacity explicit}
C=\int_1^\infty\frac{(\log (x))^2 e^{-\frac{x}{\bar{P}}}
}{\text{Ei}\left( \frac{1}{\bar{P}}\right) \bar{P}}dx\ .
\end{equation}
It is noted that Narula's approach is closely matched to the above
setting and any attempt so far to change a key parameter in this
setting (such as the entries' distribution, the number of users
per-cell, and the number of non-zero diagonals) leads to an
analytically intractable derivation. This is probably related to
the unique properties of Jacobi matrices which does not apply to
finite-band matrices in general. For example, the determinant of a
Jacobi matrix is equal to a weighted sum of the determinants of
its \emph{two} largest principal sub-matrices. In addition,
Narula's analysis provides additional evidence to support the
conjecture that the limiting spectrum of finite-band random
matrices is dependent on the distribution of their entries. On
this note, in \cite{Jing-Tse-Hou-Soriaga-Smee-Padovani-ISIT-2007}
an equivalent cellular uplink setup but with uniform phase fading
($\abs{a_{i,j}}^2=1$ and $\theta_{i,j}=\measuredangle{a_{i,j}}\sim
U[0,2\pi]$) known at the joint receiver is considered, and the
per-cell sum-rate capacity is shown to coincide with the
non-fading setup for $N\rightarrow \infty$. It is worth mentioning
that the latter result holds only for the tridiagonal case.

As an alternative to deriving exact analytical results, some works
focus on extracting parameters that characterize the channel
capacity under extreme SNR scenarios (see
\cite{Shamai-Verdu-2001-fading}\ -
\nocite{Verdu-paper-low-snr-regime-02}
\cite{Lozano-Tulino-Verdu-high-SNR-IT05} for more details on the
extreme SNR characterization). The low-SNR regime is characterized
through the minimum transmit $\ebno$ that enables reliable
communications, i.e., $\ebno_{\mathrm{min}}$, and the low-SNR
spectral efficiency slope $\So$. Assuming full receiver CSI and no
user cooperation, it is shown in
\cite{Verdu-paper-low-snr-regime-02} that the derivation of the
low-SNR parameters reduces to the calculation of
$\mysize{\trace\left(\E(\Mat{H}_N^{\dagger}\Mat{H}_N)\right)}$ and
$\mysize{\trace\left(\E\left(\Mat{H}_N^{\dagger}\Mat{H}_N\right)^2\right)}$.
For example, the low-SNR parameters for the capacity of the Wyner
setup are given for $N\rightarrow \infty$ by
\cite{Somekh-Zaidel-Shamai-CDMA-IT-CCIT-2007}
\begin{equation}\mysize\label{eq: Wyner low-snr}
\begin{aligned}
\febno_{\mathrm{min}}&=\frac{\log 2}{m_2(1+2\alpha^2)}\\
    \So&=\frac{2K(1+2\alpha^2)^2}{\mathcal{K}+K-1+4(1+K)\alpha^2+2(\mathcal{K}+2K)\alpha^4}\
    ,
\end{aligned}
\end{equation}
where the \emph{kurtosis} of an individual fading coefficient is
defined as $\mathcal{K}=m_4/(m_2)^2$. This result can be extended
in a straightforward yet tedious manner to general finite-band
matrices.

The high-SNR regime is characterized through the high-SNR slope
$\Sinf$ (also referred to as the ``multiplexing gain") and the
high-SNR power offset $\Linf$. Recently
\cite{Levy-Somekh-Shamai-Zeitouni-IT07}, the per-cell capacity
high-SNR parameters for a two diagonal $\Mat{H}_N$ ($K=1$,
$\alpha=1$, and $\beta=0$) were calculated for $N\rightarrow
\infty$ and rather general fading distributions:
\begin{equation}\mysize\label{eq: Wyner hi-snr}
\Sinf=1\quad ;\quad
\Linf=-2\max\left(\E_{\pi_a}\log_2\abs{x},\E_{\pi_b}\log_2\abs{x}\right)\
.
\end{equation}
The main idea is to link the spectral properties of
$\Mat{H}_N\Mat{H}_N^\dagger$ with the exponential growth of the
elements of its eigenvectors. Since $\Mat{H}_N\Mat{H}_N^\dagger$
in this case is an \emph{Hermitian Jacobi} matrix, and hence is
tridiagonal, its eigenvectors can be considered to be sequences
with second order linear recurrence. Therefore, the problem
reduces to the study of the exponential growth of products of two
by two matrices. This is closely related to the evaluation of the
top \emph{Lyapunov} exponent of the product; The explicit link
between the Shannon transform \eqref{eq: Mutual information} and
the top Lyapunov exponent is the \emph{Thouless} formula
\cite{groslivre}. Moreover, for arbitrary finite $K$, it is shown
in \cite{Levy-Somekh-Shamai-Zeitouni-IT07} that $\Sinf=1$ while
the power offset is bounded by a sequence of explicit upper- and
lower-bounds; the gap between the lower and the upper bounds
decreases with the bounds' order and complexity. It is noted that
calculating exact expressions for the high-SNR parameters of
channels with \emph{general} fading distribution and
\emph{arbitrary finite} $K$ remains an open problem even for the
tridiagonal case. In addition, \eqref{eq: Wyner hi-snr} also
further supports the conjecture made regarding the dependency of
the limiting spectrum of finite-band matrices on their entries'
distribution.

Recently \cite{Letzepis-PhD06}, the limiting spectrum of
$\frac{1}{1+2\alpha^2}\Mat{H}_N\Mat{H}_N^\dagger$ for the Wyner
setup and complex Gaussian vectors, has been loosely shown by
\emph{free probability} tools to be approximated by the
Mar\u{o}enko-Pastur distribution with parameter $K$. The
approximation, is shown to fairly well match the spectrum by
Monte-Carlo simulations only for relatively large values of
$\alpha$. It should be emphasized that such a match is not
guaranteed for other fading distributions excluding the complex
Gaussian distribution (i.e. Rayleigh fading). A possible reasoning
for the approximation inaccuracy in the low $\alpha$ regime is
that in the extreme case of $\alpha=0$, the eigenvalues are
evidently exponentially distributed, with no finite support (in
contrast to the Mar\u{o}enko-Pastur distribution).

\section{Concluding Remarks}
The limiting spectrum (or its Shannon transform) of certain large
finite-band Hermitian random matrices is known for a few limited
cases and remains an open problem in general. Moreover, even the
high-SNR characterization of their Shannon transforms is still
unsolved. Due to their special power profile, standard tools from
the theory of random matrices cannot be used for this problem. It
is conjectured that unlike ``full" random matrices, the limiting
spectra of finite-band random matrices depend on the actual
distribution of their entries. It seems that unconventional
methods such as the method used by Narula, replacing the role of
eigenvalues with the diagonal elements of the Cholesky
decomposition, are required to shed light on this problem.
Nevertheless, it is noted that the tri-diagonal (Jacobi matrices)
case is unique and these techniques may not apply to general
finite-band matrices.
Finally, we note that solving the problem would facilitate
analytical treatment, which in turn gains much insight into the
effect of key system parameters on the performance of certain
cellular uplink channels and time varying ISI channels.

\section*{Acknowledgment}

The research was supported in part by a Marie Curie Outgoing
International Fellowship and the NEWCOM++ network of excellence
both within the 6th European Community Framework Programme, by the
U.S. National Science Foundation under Grants CNS-06-25637 and
CNS-06-26611, and also by the REMON Consortium.

\bibliographystyle{ieeetr}


\end{document}